\documentclass[12pt,aps,pre,preprint,amssymb,groupedaddress,showpacs]{revtex4}

\usepackage{graphicx}
\usepackage{epsfig}
\def\sgn{\mathop{\textrm{sgn}}\nolimits}
\def\sech{\mathop{\textrm{sech}}\nolimits}
\def\tanh{\mathop{\textrm{tanh}}\nolimits}
\def\fr{\mbox{$\frac{1}{2}$}}

\begin{document}

\title{Transverse instability and its long-term development
  for solitary waves of the (2+1)-Boussinesq equation}

\author{K.B. Blyuss}
\author{T.J. Bridges}
\author{G. Derks}
\affiliation{Department of Mathematics \& Statistics,
University of Surrey, Guildford GU2 7XH, UK}  

\date{\today}

\begin{abstract}
The stability properties of line solitary wave solutions of the
(2+1)-dimensional Boussinesq equation with respect to transverse
perturbations and their consequences are considered. A geometric
condition arising from a multi-symplectic formulation of this equation
gives an explicit relation between the parameters for transverse
instability when the transverse wavenumber is small.
The Evans function is then computed explicitly,
giving the eigenvalues for transverse instability for all transverse
wavenumbers.  To determine the nonlinear and long time implications of
transverse instability, numerical
simulations are performed using pseudospectral discretization.  The
numerics confirm the analytic results, and in all cases studied,
transverse instability leads to collapse.
\end{abstract}


\maketitle

\section{Introduction}

One of the fundamental ways that a solitary wave traveling in one
space dimension generates a two space dimensional pattern is through
transverse instability.  A transverse instability of a line solitary
wave is associated with a class of perturbations traveling in a
direction transverse to the basic direction of
propagation.  In addition to establishing the existence of
transverse instability, a major question is what implications this
instability means for the long-term behaviour of the system: does it
settle into a new two-space-dimensional pattern, or collapse ?  In
this paper we study this sequence of questions for the canonical
Boussinesq equation in two space dimensions
\begin{equation}\label{1}
u_{t t}=(f(u)+\varepsilon u_{x x})_{x x}+\sigma u_{y y}\,,
\end{equation}
where $\varepsilon=\pm1$
and $\sigma=\pm1$.  In general, $f(u)$ can be any smooth function, but
the canonical form of the Boussinesq equation has the form
\[
f(u) =D(u^2-u)\quad\mbox{with}\quad D=\pm 1\,.
\]

When $D=-1$, $\varepsilon=1$ and $\sigma=1$ this equation was derived
by Johnson \cite{J} to describe the
propagation of gravity waves on the surface of water, in particular
the head-on collision of oblique waves, and it was derived by
Breizman and Malkin \cite{BM} in the context of Langmuir waves.

In the absence of the
transverse variation (i.e, $u_{y}=0$) and for $\varepsilon=-1$, $D=-1$
this equation reduces to the so-called "good"
Boussinesq equation, which is well-posed, and for which
$\sech^{2}$-solutions exist for any $c$ with $|c|<1$. These waves are
stable when $\frac{1}{2}<|c|<1$ \cite{BS}.  For the case
$|c|<\frac{1}{2}$ it was shown by computer-assisted simulation of the
leading term in the Taylor expansion of the Evans function that there
is an unstable eigenvalue \cite{AS}.  This result was 
generalized to include solitary waves with nonzero tails, and
rigorously proved using the symplectic Evans matrix in \cite{BD1}.

Transverse instability of solitary waves has been widely studied
since the seminal work of Zakharov \cite{VEZ} on the nonlinear
Schr\"odinger equation and the work of Kadomtsev \& Petviashvili
\cite{KP} on transverse instability of the Korteweg-de Vries soliton. 
Since then, transverse instability of solitary waves has been
investigated for a wide range of models; examples include the nonlinear
Schr\"odinger (NLS) equation and related equations \cite{KRZ,LS,WAS},
Kadomtsev-Petviashvili equation \cite{APS,ARa,IRS,KPe}, the
Zakharov-Kuznetsov equation \cite{AR,B1,LS}, and water waves \cite{B3}.
A review of transverse
instability for NLS and other related models can be found in
Kivshar \& Pelinovsky \cite{KPe}.

In this paper, we will first use a geometric condition as derived in
\cite{B1} to get an explicit criterion for small transverse wavenumber
instability.
For this we use the multi-symplectic formulation of (\ref{1}) in an
essential way.
To get detailed information for all transverse wavenumbers we compute
explicitly the Evans function for the (2+1)-dimensional Boussinesq model
linearized about a larger family of line solitary waves (allowing
the state at infinity to be nonzero).  Plots of the dependence of the
growth rate on the transverse wavenumber are presented. 

The post-instability behaviour of the nonlinear problem is studied
using direct numerical simulation.  The numerical evidence confirms
the analytic results and suggests that the post-instability in the nonlinear
system leads to collapse in all cases.
A multi-symplectic pseudospectral discretization \cite{BR} is
used as a basis for the numerical simulations.
The numerical scheme is applied to the full
two-dimensional PDE and we observe transverse modulation and further
development of the longitudinal and transverse instabilities,
resulting in the collapse of the initial line solitary waves. 
In the parameter region where the analytic criterion indicates that
the solitary wave state is longitudinally stable but transversely
unstable, simulations support the analytic results and provide insight into
the long-term development of this instability.

\section{Multi-symplectifying the equations}

The Boussinesq system has a range of geometric structures.  Firstly,
we record the Lagrangian and Hamiltonian structures.  Let
$u=\phi_{xx}$, then the system is Lagrangian with
\[
L =\int\left[ -\fr\phi_{xt}^2 +F(\phi_{xx}) + \fr\varepsilon\phi_{xxx}^2
+\fr\sigma\phi_{xy}^2\,\right]dxdydt,
\]
where $F(\cdot)$ is any function satisfying $F'(\cdot)=f(\cdot)$.

The Boussinesq equation can be represented as a Hamiltonian system
in a number of ways (e.g. \cite{T}).  For example, let
\[
H=\int\left[F(u)-\fr\varepsilon u_{x}^{2}+\fr\Phi_{x}^{2}
+\fr\sigma w_{y}^{2} +\gamma(u-w_x)\right]\,dxdy\,,
\]
where $\gamma$ is a Lagrange multiplier associated with the
constraint $u=w_x$.  With Hamiltonian variables $(\Phi,u,w,\gamma)$ the
governing equations take the form
\begin{equation}
\begin{array}{lll}
-u_{t} &=&\frac{\delta H}{\delta\Phi}= -\Phi_{x x}\\
\Phi_{t}&=&\frac{\delta H}{\delta u}=f(u)+\varepsilon u_{x
  x}+\gamma\,,\\
0 &=& \frac{\delta H}{\delta w} = \gamma_x - \sigma w_{yy}\,,\\
0 &=& \frac{\delta H}{\delta \gamma} = u-w_x\,.
\end{array}
\end{equation}

However, the most interesting form of (\ref{1}) for the present
purposes is the multi-symplectic formulation which can be represented
in the canonical form \cite{BD3}
\begin{equation}\label{3}
{\bf M}Z_{t}+{\bf K}Z_{x}+{\bf L}Z_{y}=\nabla S(Z)\,
\quad Z\in{\mathbb R}^6\,,
\end{equation}
where
\[
Z=\left(
\begin{array}{c}
q_{1}\\
q_{2}\\
q_{3}\\
p_{1}\\
p_{2}\\
p_{3}\\
\end{array}
\right),\mbox{ }
{\bf M}=\left(
\begin{array}{cccccc}
0&1&0&0&0&0\\
-1&0&0&0&0&0\\
0&0&0&0&0&0\\
0&0&0&0&0&0\\
0&0&0&0&0&0\\
0&0&0&0&0&0\\
\end{array}
\right),\quad \mbox{with}\quad u(x,y,t) = q_1(x,y,t)\,,
\]
\[
{\bf K}=\left(
\begin{array}{cccccc}
0&0&0&1&0&0\\
0&0&0&0&1&0\\
0&0&0&0&0&1\\
-1&0&0&0&0&0\\
0&-1&0&0&0&0\\
0&0&-1&0&0&0\\
\end{array}
\right),\mbox{ }
{\bf L}=\left(
\begin{array}{cccccc}
0&0&1&0&0&0\\
0&0&0&0&0&0\\
-1&0&0&0&0&0\\
0&0&0&0&0&0\\
0&0&0&0&0&0\\
0&0&0&0&0&0\\
\end{array}
\right),
\]
\[
S(Z)=-F(q_{1})-\frac{1}{2\varepsilon}p_{1}^{2}+\frac{1}{2}p_{2}^{2}-\frac{\sigma}{2}p_{3}^{2}.
\]
Using $q_1=u$ it is straightforward to show that
this system is a reformulation of (\ref{1}).

\section{Geometric criterion for transverse instability}

An advantage of the multi-symplectic formulation is that
there is a geometric condition which is easy to verify for
transverse instability of line solitary waves \cite{B1}.

Consider the well-known basic family of solitary waves
of (\ref{1}) of the form
\begin{equation}\label{4}
Z(x,y,t)=\widehat{Z}(\theta;c,l),\mbox{ }\theta=x-ct+l y +\theta_0,
\end{equation}
obtained by taking the first component to be a ${\rm sech}^2$ wave,
\begin{equation}\label{sw}
u(\theta;c,l) = \langle {\bf e}_1,\widehat Z(\theta;c,l)\rangle
=A(c,l)\,\sech^2(B(c,l)\theta)\,,
\end{equation}
with
\[
B(c,l) = \fr\sqrt{ \varepsilon(D+c^2-\sigma l^2)}\,,\quad
A(c,l) = 6\frac{\varepsilon}{D}B^2\,.
\]
Existence of the solitary wave clearly
requires $\varepsilon(D+c^2-\sigma l^2)>0$.
The other components of $\widehat Z$ are easily obtained from
(\ref{sw}) and the multi-symplectic equations (\ref{3}).

For the linear stability analysis, let
$Z(x,y,t)=\widehat{Z}(\theta;c,l)+\Re[U(\theta;\lambda,k)e^{\lambda
  t+iky}]$, substitute this into (\ref{3}) and linearize. Then, if the
  resulting linear equation has square-integrable solutions
  $U(\theta;\lambda,k)$ with $\Re(\lambda)>0$ and $k\in{\mathbb R}$,
  we call the basic solitary wave state $\widehat{Z}(\theta;c,l)$ transversely
  unstable. Assuming that $\widehat{Z}_{\theta}$ is the only square
  integrable element in the kernel of the linearization operator
  $\mathcal{L}=D^{2}S(Z)-\left[{\bf K}-c{\bf M}+l{\bf
  L}\right]\frac{d}{d\theta}$, we have the following geometric
  condition of transverse instability for small $\lambda$ and
  $k$. Suppose
\begin{equation}\label{7}
\Delta=
\left|
\begin{array}{cc}
{\mathcal A}_{c}&{\mathcal A}_{l}\\
{\mathcal B}_{c}&{\mathcal B}_{l}\\
\end{array}
\right|>0,
\quad \mbox{where}\quad\left\{\,
\begin{array}{l}
{\mathcal A}=-\frac{1}{2}\int_{-\infty}^{\infty}\langle{\bf
  M}\widehat{Z}_{\theta},\widehat{Z}\rangle d\theta,\\
{\mathcal B}=\frac{1}{2}\int_{-\infty}^{\infty}\langle{\bf
  L}\widehat{Z}_{\theta},\widehat{Z}\rangle d\theta\,,\\
\end{array}\right.
\end{equation}
Then the basic solitary wave $\widehat{Z}(\theta;c,l)$ 
of (\ref{1}) is linearly transverse unstable \cite{B1,B3}.

Using the above definitions of the multi-symplectic
 matrices ${\bf M}$ and ${\bf L}$, we obtain
\begin{equation}\label{9}
{\mathcal A}=-\frac{1}{2}\int_{-\infty}^{\infty}\left(q_{1}\frac{d}{d\theta}q_{2}-q_{2}\frac{d}{d\theta}q_{1}\right)d\theta=-c\int_{-\infty}^{\infty}q_{1}^{2}d\theta=-cK,
\end{equation}
\begin{equation}\label{10}
{\mathcal
  B}=\frac{1}{2}\int_{-\infty}^{\infty}\left(q_{1}\frac{d}{d\theta}q_{3}-q_{3}\frac{d}{d\theta}q_{1}\right)d\theta=\sigma l\int_{-\infty}^{\infty}q_{1}^{2}d\theta={\sigma l}K,
\end{equation}
where
\[
K=\int_{-\infty}^{\infty}u^{2}d\theta=\frac{4}{3}\frac{A^{2}}{B}=-\frac{6\varepsilon}{D^{2}}\left(\sigma
    l^{2}-c^{2}-D\right)\sqrt{-\frac{\sigma
    l^{2}-c^{2}-D}{\varepsilon}}.
\]
Substitution of (\ref{9}) and (\ref{10}) in (\ref{7}) yields:
\[
\sgn\Delta=\sgn\left[-\frac{\sigma}{c}{\mathcal A}\left({\mathcal
      A}_{c}+\frac{l}{c}{\mathcal
      A}_{l}\right)\right]=\sgn\left[\sigma\left({\mathcal A}_{c}+\frac{l}{c}{\mathcal A}_{l}\right)\right]=
\]
\begin{equation}
=\sgn\left[-\sigma\left(K+c\frac{\partial}{\partial
  c}K+l\frac{\partial}{\partial l}K\right)\right]=\sgn\left[-\sigma\left(\sigma
  l^{2}-c^{2}-D\right)\left(4\sigma l^{2}-4c^{2}-D\right)\right].
\end{equation}
Since the condition for transverse instability requires $\Delta>0$, we
have the following result: {\it Suppose
\begin{equation}\label{12}
\varepsilon\sigma\left(4\sigma l^{2}-4c^{2}-D\right)>0,
\end{equation}
then the basic solitary wave $\widehat{Z}(\theta;c,l)$ is linearly
transversely unstable}.

The multi-symplectic formulation also provides an expression for the
linear growth rate of the instability $\lambda$ as a function of the
transverse wavenumber $k$ for long-wave perturbations \cite{B1}:
\begin{equation}
\lambda=\frac{\sqrt{{\mathcal A}_{c}{\mathcal
    B}_{l}-{\mathcal A}_{l}{\mathcal
    B}_{c}}}{|{\mathcal A}_{c}|}k+\mathcal{O}(k^{2})=\frac{\sqrt{-\sigma(4\sigma l^2-4c^2+1)(\sigma l^2-c^2+1)}}{4c^2-1-\sigma l^2}k+\mathcal{O}(k^{2}).
\end{equation}
This provides the growth rate for $k$ small.  In the next section, the
Evans function will be constructed in order to determine the growth
rate for all transverse wavenumbers $k$.

In the remainder of this section, we apply the condition (\ref{12})
for various parameter values.

For the "good" Boussinesq equation from
\cite{BS} with $\varepsilon=-1$ and $D=-1$ the existence and
transverse instability requirements are
\begin{equation}
\sigma l^{2}-c^{2}+1>0\quad\mbox{and}\quad
-\sigma\left(\sigma l^{2}-c^{2}+\frac{1}{4}\right)>0.
\end{equation}
respectively.  Combining these conditions
leads to the following system of
inequalities for $c$ and $l$ when $\sigma>0$
\begin{equation}\label{16}
\frac{1}{4}+\sigma
l^{2}<c^{2}<1+\sigma l^{2},
\end{equation}
and for $\sigma<0$
\begin{equation}\label{17}
c^{2}<\frac{1}{4}+\sigma l^{2}
\end{equation}
These inequalities define the regions in $(c,l)$ parameter plane,
where the basic solitary wave exits and is
linearly transversely unstable, and these regions are
presented in Figure 1.

One can do a similar analysis for Johnson's equation \cite{J},
where $\sigma=1$, $\varepsilon=1$ and $D=-1$.
The existence requirement is $l^{2}<c^{2}-1$ and the instability
condition is $l^{2}>c^{2}-\frac{1}{4}$.  This result is inconclusive
for two reasons.  First, the two regions do not overlap so
the geometric condition does not predict instability for any parameter
values.  Secondly, when $\varepsilon=+1$ the equation is ill-posed as
an evolution equation (this can be seen at the linear level where the
dispersion relation predicts instability as the wavenumber goes to
infinity), and so the question of long time stability is irrelevant.
\begin{figure}
\epsfig{width=5cm,file=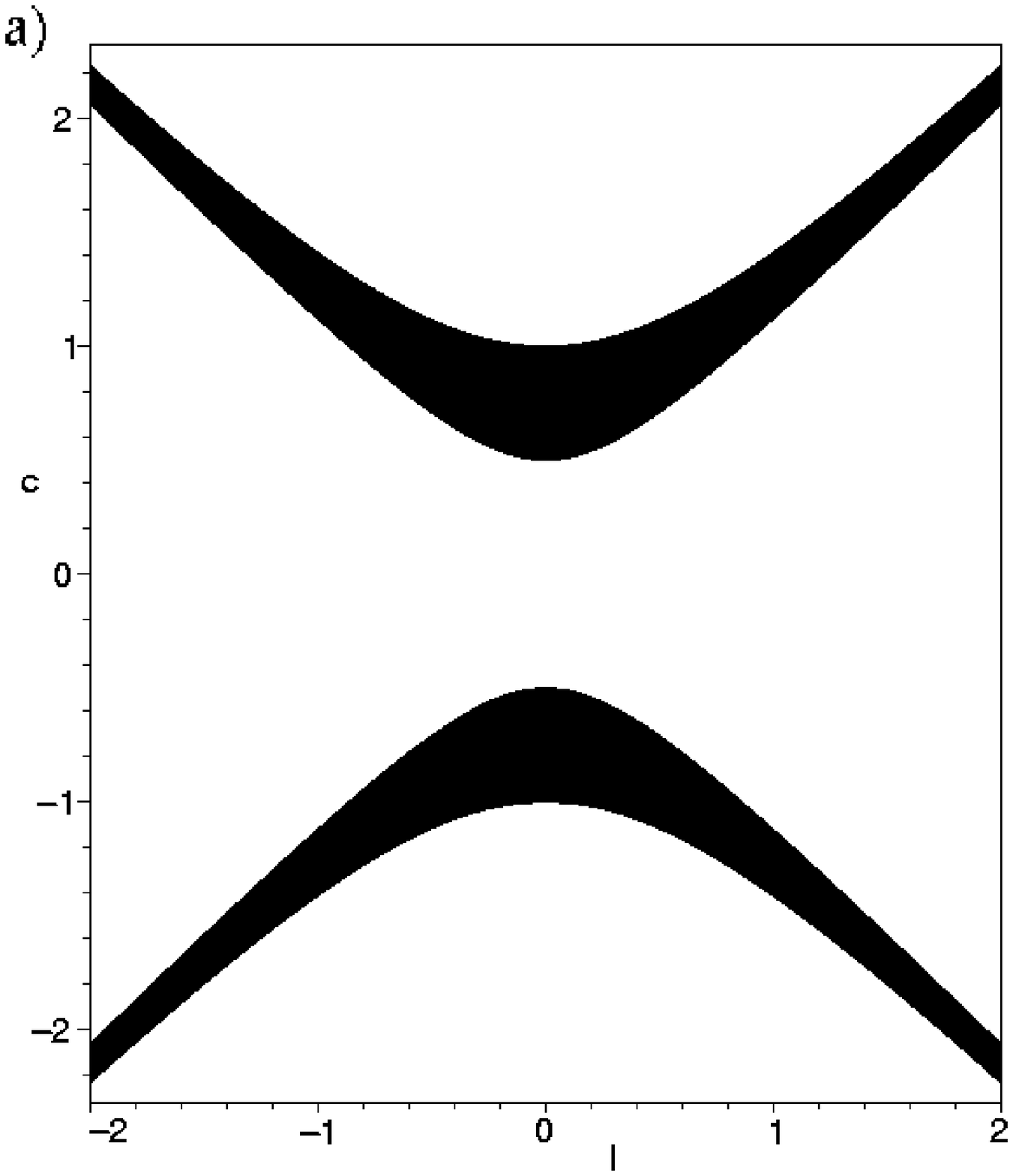}
\hspace{1cm}
\epsfig{width=5.6cm,file=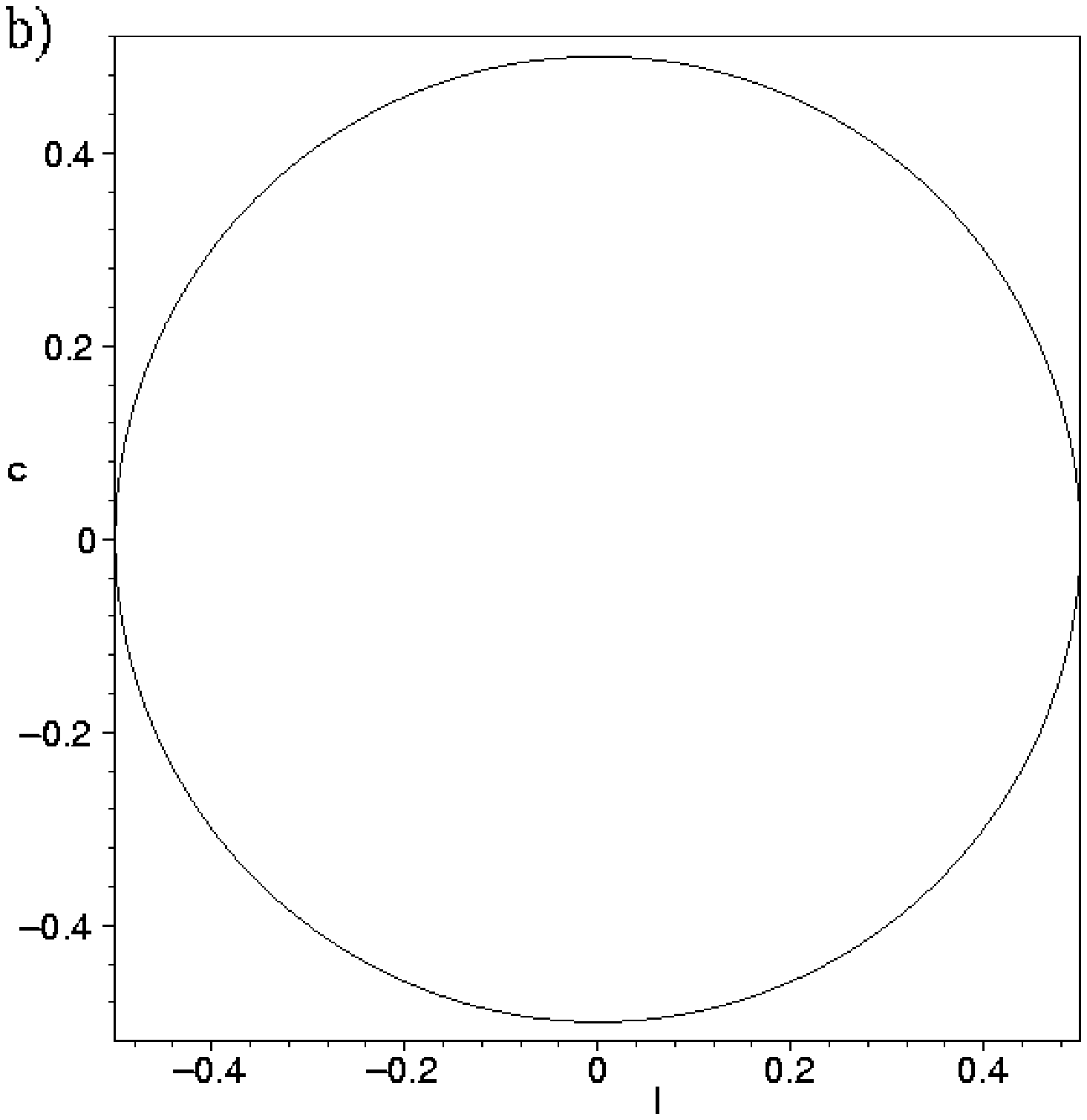}
\caption{Theoretical boundaries of transverse instability. a) Case
  (\ref{16}) with $\sigma=1$. The waves are unstable for the
  parameters lying within the shaded regions. b) Case
  ({\ref{17}}) with $\sigma=-1$. The waves are unstable for the
  parameters within the circle.}
\end{figure}

\section{Evans function analysis of the transverse instability}

In this section we use the Evans function formalism in order to
analyze the linear transverse stability problem for the Boussinesq model
(\ref{1}) for all values of the transverse wavenumber.  We restrict
attention to the parameter values of most interest:
$\varepsilon=-1$ and $D=-1$ associated with the ``good'' Boussinesq,
although we put no restriction on $\sigma$ (but keeping in mind that
$\sigma=+1$ is the most interesting case).

However, the class of solitary waves will be enlarged. Namely, we
include solitary waves bi-asymptotic to a nontrivial state at infinity,
specifically,
\begin{equation}\label{24}
U(\theta)=U_\infty + 6\delta^{2}\sech^{2}\left(\delta\theta\right),
\quad\theta=x-ct+ly,
\end{equation}
where
\begin{equation}\label{25}
\delta=\frac{1}{2}\sqrt{\sqrt{1+4a}-c^{2}+\sigma l^2}\quad\mbox{and}\quad
U_\infty = \fr(1-c^2+\sigma l^2)-2\delta^2=-\frac{2a}{1+\sqrt{1+4a}}\,.
\end{equation}
The value of the
parameter $a$ is constrained only by existence of the square
root: $1+4a\geq (c^2-\sigma l^2)^2$.

Here we will not use any geometric structure (although it might
be interesting to look more closely in this direction) and so
work directly with (\ref{1}).  Let
\begin{equation}\label{26}
u(x,y,t)=U(\theta)+\Re{\left(\tilde{u}(\theta)\exp\left[iky+\lambda
      t\right]\right)}.
\end{equation}
By substituting this expression in (\ref{1}) and linearising, 
one obtains the following equation for the complex function 
$\tilde{u}(\theta)$
\begin{equation}\label{27}
\tilde{u}_{\theta\theta\theta\theta}+2(U\tilde{u})_{\theta\theta}-(1-c^{2}+\sigma
l^2)\tilde{u}_{\theta\theta}-2(c\lambda+i\sigma kl)
\tilde{u}_{\theta}+(\lambda^{2}+\sigma k^{2})\tilde{u}=0.
\end{equation}
After the change of variable $\tilde{x}=\delta\theta$, substitution of
the explicit expression for $U$ from (\ref{24}), and dropping the
tildes, equation (\ref{27}) reduces to
\begin{equation}\label{28}
u_{x x x x}-4\left[(1-3\sech^{2}x)u\right]_{x x}-\gamma u_{x}+\beta u=0,
\end{equation}
where
\begin{equation}\label{29}
\gamma=\frac{2(c\lambda+i\sigma kl)}{\delta^{3}}\mbox{ and
  }\beta=\frac{\lambda^{2}+\sigma k^{2}}{\delta^{4}}.
\end{equation}
To obtain explicit solutions of this equation, we note that
by taking $u=\phi_{x x}$ and $v=1-3\sech^{2}x$ in (\ref{28}), and
integrating twice the equation simplifies to
\begin{equation}\label{31}
\phi_{x x x x}-4v\phi_{x x}-\gamma\phi_{x}+\beta\phi=0.
\end{equation}
Solutions of this equation can be readily found in a manner similar to
that in \cite{B} (see also \cite{APS}). First we note that in the
limit $x\to\pm\infty$, equation (\ref{31}) reduces to
\begin{equation}
\phi_{x x x x}-4\phi_{x x}-\gamma\phi_{x}+\beta\phi=0.
\end{equation}
Substituting now $\phi=e^{\mu x}\hat{\phi}$, one can see that $\mu$
satisfies the quartic equation
\begin{equation}\label{33}
\mu^{4}-4\mu^{2}-\gamma\mu+\beta=0\,.
\end{equation}
Quartics of this form have been analyzed in \cite{BD1} (see equation
(10.9) there), and when $\Re(\beta)>0$
there are two roots with positive real
part and two roots with negative real part.  Therefore, the space of
solutions decaying as $x\to +\infty$ is two-dimensional, as is the the
space of solutions decaying as $x\to -\infty$.  

If the four roots
$\mu_{j},\mbox{ }j=1,..,4$ of the equation (\ref{33}) are distinct, the
corresponding solutions of (\ref{31}) are given by
\begin{equation}\label{34}
\phi_{j}(x)=e^{\mu_{j}x}h_{j}(x),
\end{equation}
with
\begin{equation}\label{35}
h_{j}(x)=(4\mu_{j}^{3}+8\mu_{j}-\gamma)-12\mu_{j}^{2}\tanh x.
\end{equation}
The case of multiple roots can be handled similarly \cite{B}.
Solutions of the original equation (\ref{28}) are found by
substituting $u(x)=\phi(x)_{x x}$, and the other components of the vector
${\bf v}(x)$ can be obtained by the differentiating the expression for
$u(x)$.

Localised solutions of the linearised problem exist if one can match
the solutions decaying as $x\to\infty$ with the solutions decaying as
$x\to -\infty$. This can be determined by finding the zeros of the so-called
Evans function which correspond to the eigenvalues of the linearised
problem. To define the Evans function, we write the equation
(\ref{28}) as a first-order system
\begin{equation}\label{30}
{\bf v}_{x}=A(x){\bf v},\mbox{  }{\bf v}=\left(
\begin{array}{l}
u\\
u_{x}\\
u_{x x}\\
u_{x x x}
\end{array}\right),\mbox{  }A(x)=\left(
\begin{array}{cccc}
0&1&0&0\\
0&0&1&0\\
0&0&0&1\\
-\beta+4v_{x x}&\gamma+8v_{x}&4v&0
\end{array}
\right)
\end{equation}
with $v=1-3\sech^{2}x$.

Since the trace of the matrix $A(x)$ vanishes, the Evans function
can be defined as $E(\lambda,k)={\bf v}_{1}(x)\wedge{\bf
  v}_{2}(x)\wedge{\bf v}_{3}(x)\wedge{\bf v}_{4}(x)$ \cite{AGJ}. An
alternative expression for the Evans function can be derived by using
the adjoint system as shown in \cite{BD2}. The adjoint system of
(\ref{30}) has the form:
\begin{equation}\label{36}
{\bf w}_{x}=-A(x)^{*}{\bf w},\mbox{  }{\bf w}=\left(
\begin{array}{l}
w_{1}\\
w_{2}\\
w_{3}\\
w_{4}
\end{array}\right),
\end{equation}
where $A(x)^{*}$ denotes the Hermitian conjugate of $A$ $(A(x)^{*}=\overline{A(x)}^{T})$.
The equation for $w_{4}$ turns out to be
\begin{equation}
(w_{4})_{x x x x}-4v(w_{4})_{x x}+\overline{\gamma}(w_{4})_{x}+\overline{\beta}(w_{4})=0.
\end{equation}
This equation is equivalent to (\ref{31}) up to the change of
variables: $x\to -x$, $\gamma\to\overline{\gamma}$,
$\beta\to\overline{\beta}$, and therefore its solutions can be
obtained from (\ref{34}) by changing $x$ for $-x$ and
conjugating them:
\begin{equation}
(w_{4})_{j}=e^{-\mu_{j}^{*}x}\overline{h_{j}(-x)},
\end{equation}
with $h_{j}(x)$ defined in (\ref{35}). Other components of the vector
${\bf w}(x)$ can be obtained from (\ref{36}).

Let $\mu_{1}$ and $\mu_{2}$ be the two roots of the equation (\ref{33})
with negative real part, and let ${\bf v}_{j}(x)$ and ${\bf
  w}_{j}(x)$, $j=1,2$ be the corresponding solution vectors of the
linearised (respectively, adjoint) system. Since the matrix
$A(x)$ in (\ref{30}) is traceless, we can define the Evans function
for the system (\ref{30}) as follows \cite{BD2}:
\begin{equation}\label{39}
E(\lambda,k)=\left|
\begin{array}{cc}
\langle {\bf w}_{1}(0),{\bf v}_{1}(0)\rangle & \langle {\bf w}_{1}(0),{\bf
  v}_{2}(0)\rangle\\
\langle {\bf w}_{2}(0),{\bf v}_{1}(0)\rangle & \langle {\bf w}_{2}(0),{\bf
  v}_{2}(0)\rangle
\end{array}
\right|,
\end{equation}
where $\langle\cdot,\cdot\rangle$ denotes the complex inner
product in $\mathbb{C}^{4}$. To obtain a unique definition of the
Evans function, the scaling
$\lim_{x\to\infty}e^{-2\mu_{j}x}\langle{\bf w}_{j}(-x),{\bf
  v}_{j}(x)\rangle=1$ is used. This normalises the eigenvectors and
the adjoint eigenvectors of $A^{\infty}=\lim_{x\to\pm\infty}A(x)$.

\begin{figure}
\epsfig{width=7cm,file=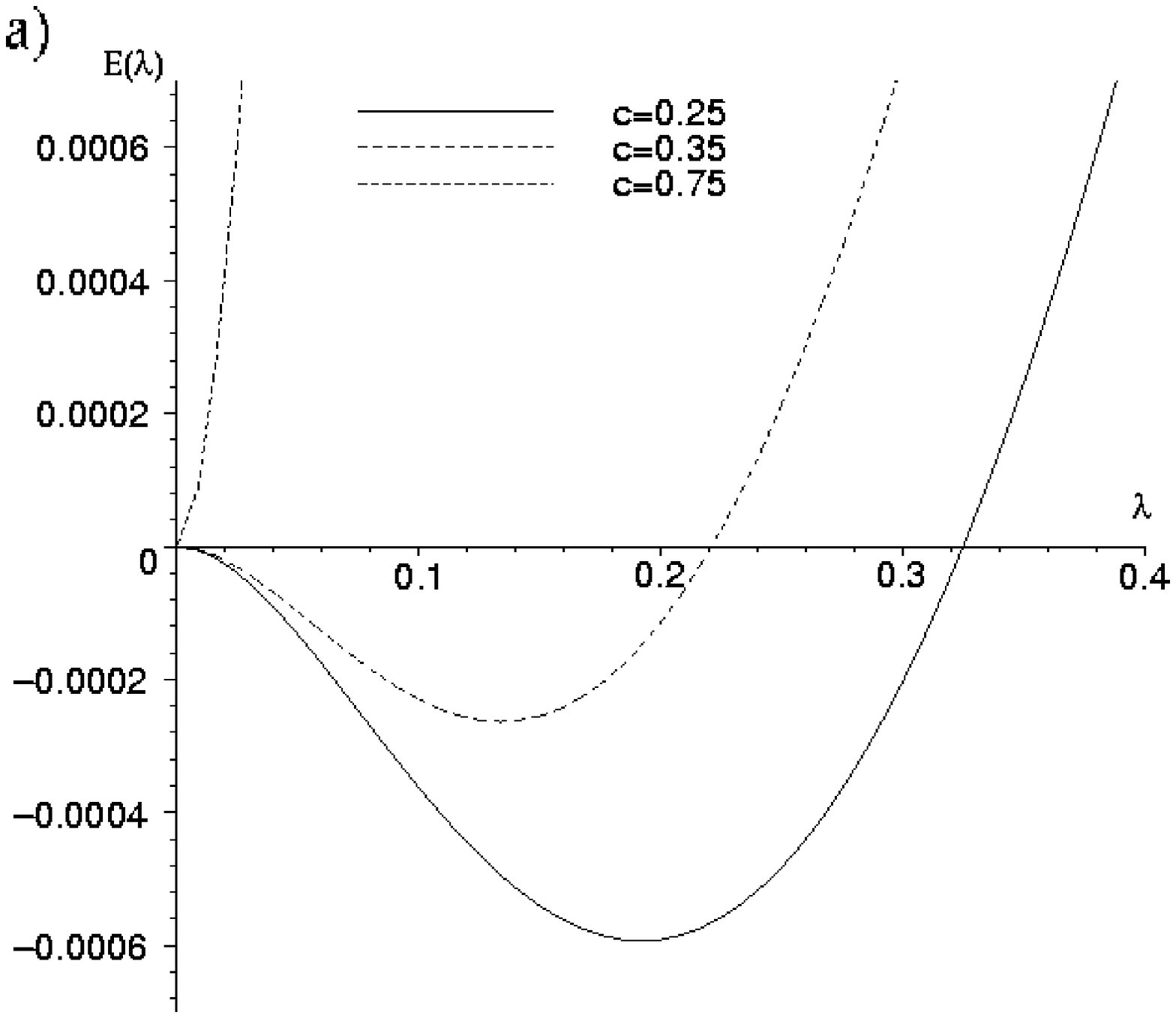}
\hspace{0.6cm}
\epsfig{width=4.6cm,file=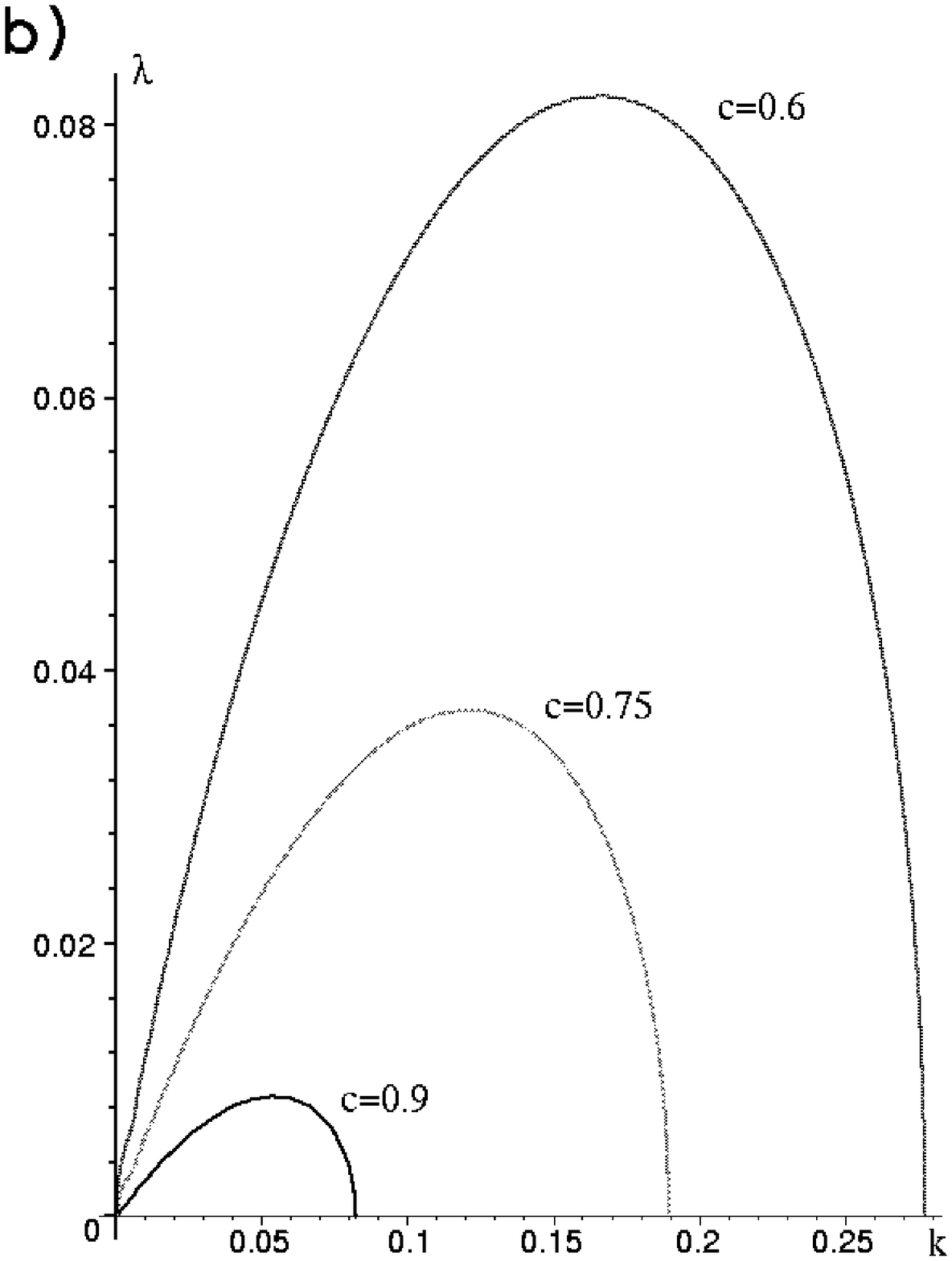}
\caption{a) The Evans function $E(\lambda)=E(\lambda,0)$ versus
  $\lambda$ for $c=0.25$, $c=0.35$ and $c=0.75$ respectively. b)
  Growth rate versus transverse wavenumber for the values of velocity
  $c=0.6$, $c=0.75$ and $c=0.9$ respectively.}
\end{figure}

After some lengthy algebra and
introducing the scaling, which enforces the
asymptotic limit $E(\lambda,k)\to 1$ as
$\lambda\to\infty$, the final expression for the Evans function can be
obtained, which we do not present here since it is lengthy (the
expression for the Evans function as well as the calculations of the
instability growth rate can be downloaded as a
\textsc{Maple}-file from the website \cite{Web}).  

Zeros of the Evans function $E(\lambda,k)$ correspond to the bounded solutions
of the linearised stability problem with the wavenumber $k$ and the growth rate
$\Re(\lambda)$.  The leading order terms (in $k$ and $\lambda$) in the
Evans function are in complete agreement with the results of the geometric
condition of \S3. Note that, since the construction here is based
on a basic solitary wave with a nontrivial state at infinity, it is
suggestive that the geometric condition \cite{B1} extends to such waves.

We illustrate the dependence of the Evans function on the wavespeed
and transverse wavenumber
in Figure 2.  In the left graph, the transverse wavenumber
is set to zero, to compare with known results on longitudinal
instability.  The graph is in complete agreement with known
results (e.g. \cite{BS,BD1}) that the solitary wave is stable for
$\frac{1}{2}<c\leq 1$ and unstable for $0\leq c<\frac{1}{2}$.  In the
right-hand graph in Figure 2 we present the plot of the growth
rate $\Re(\lambda)$ as a function of the transverse wavenumber. 
Note that waves of the good Boussinesq which are longitudinally stable
are transverse unstable.  Note also that there is a cut-off
wavenumber, similar to other cases of transverse instability, such as
in the Zakharov-Kuznetsov equation \cite{AR}.

\section{Post-instability simulations}

In this section we perform a simulation of the PDE (\ref{1}) using
the multi-symplectic spectral discretization proposed in \cite{BR} and
applied there to Zakharov-Kuznetsov and shallow-water equations.

The (2+1)-dimensional Boussinesq equation is considered with
$\varepsilon=D=-1$ on a finite domain $(x,y)=[0,L]\times[0,L]$ with $L>0$
some constant, and periodic boundary conditions on both spatial
variables. We choose a spatial mesh-size as $\Delta x=\Delta
y\equiv\Delta m=L/2N$ and introduce the discrete two-dimensional
Fourier transform defined as
\[
U_{k l}=\frac{1}{\sqrt{2N}}\sum_{i,j=1}^{2N}u_{i j}e^{-\theta_{k}(i-1)\Delta
  m-\theta_{l}(l-1)\Delta m},
\]
where
\[
\theta_{k}=i\frac{2\pi (k-1)}{L}
\]

and $u_{i j}\approx u(m_{i j}),m_{i j}=(i-1)\Delta x+(j-1)\Delta y$
(cf. \cite{F}). Fourier spectral discretization of the
(2+1)-dimensional Boussinesq equation yields
\begin{equation}
\partial_{t t}U_{k
  l}=\bar{\theta}_{k}^{2}\left[\varepsilon\bar{\theta}_{k}^{2}U_{k
  l}+\nabla_{k l}\bar{F}({\bf
  U})\right]+\sigma\bar{\theta}_{l}^{2}U_{k l},
\end{equation}
where $\bar{\theta}_{k}$ are the entries of the diagonal matrix
defined by the relations
\[
\bar{\theta}_{k}=\theta_{k},\mbox{ for }k=1,...,N,
\]
$\bar{\theta}_{N+1}=0$, and
\[
\bar{\theta}_{k}=-\theta_{2N-k+2},\mbox{ for }k=N+2,...,2N
\]
\begin{figure}
\epsfig{width=5cm,file=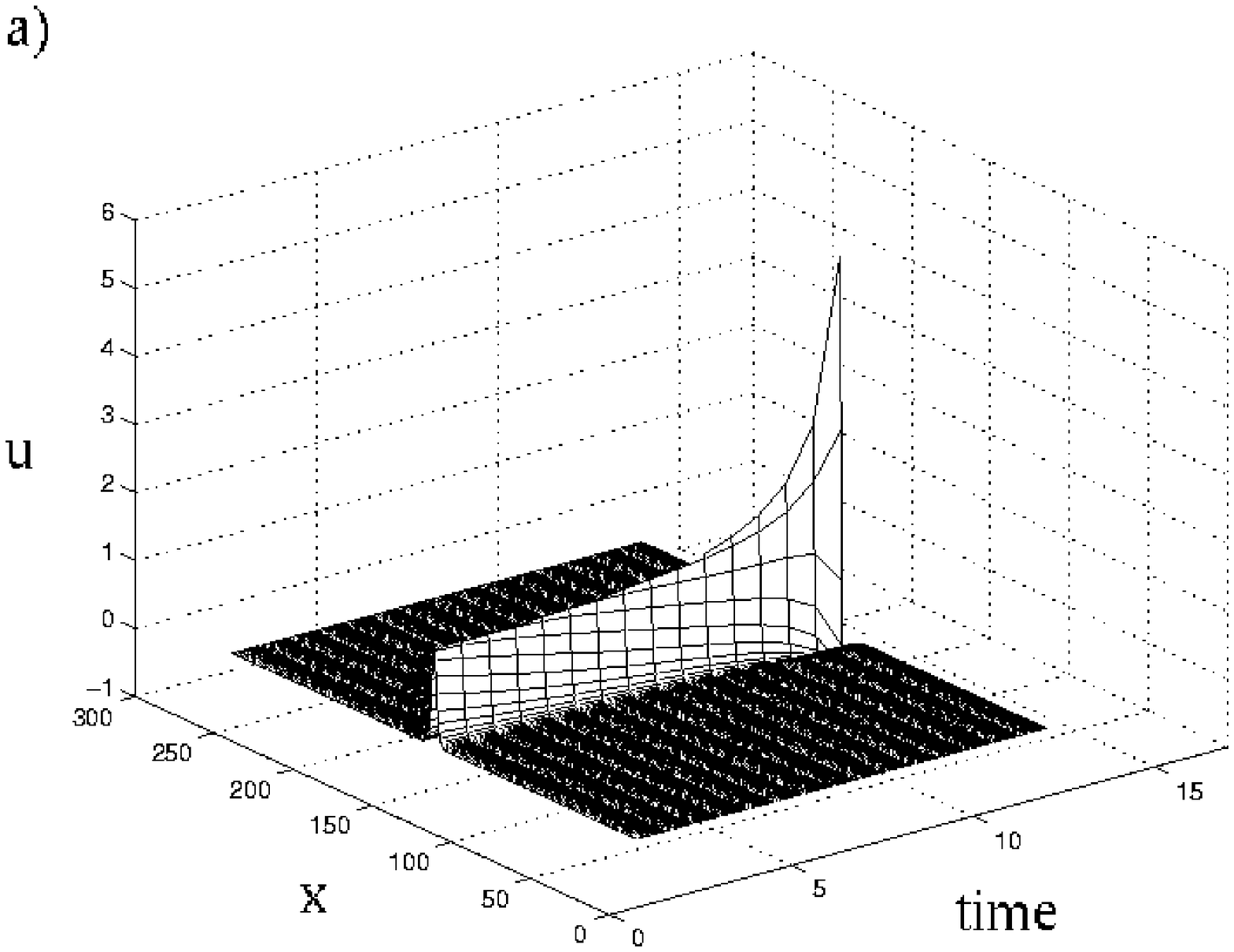}
\hspace{1cm}
\epsfig{width=5cm,file=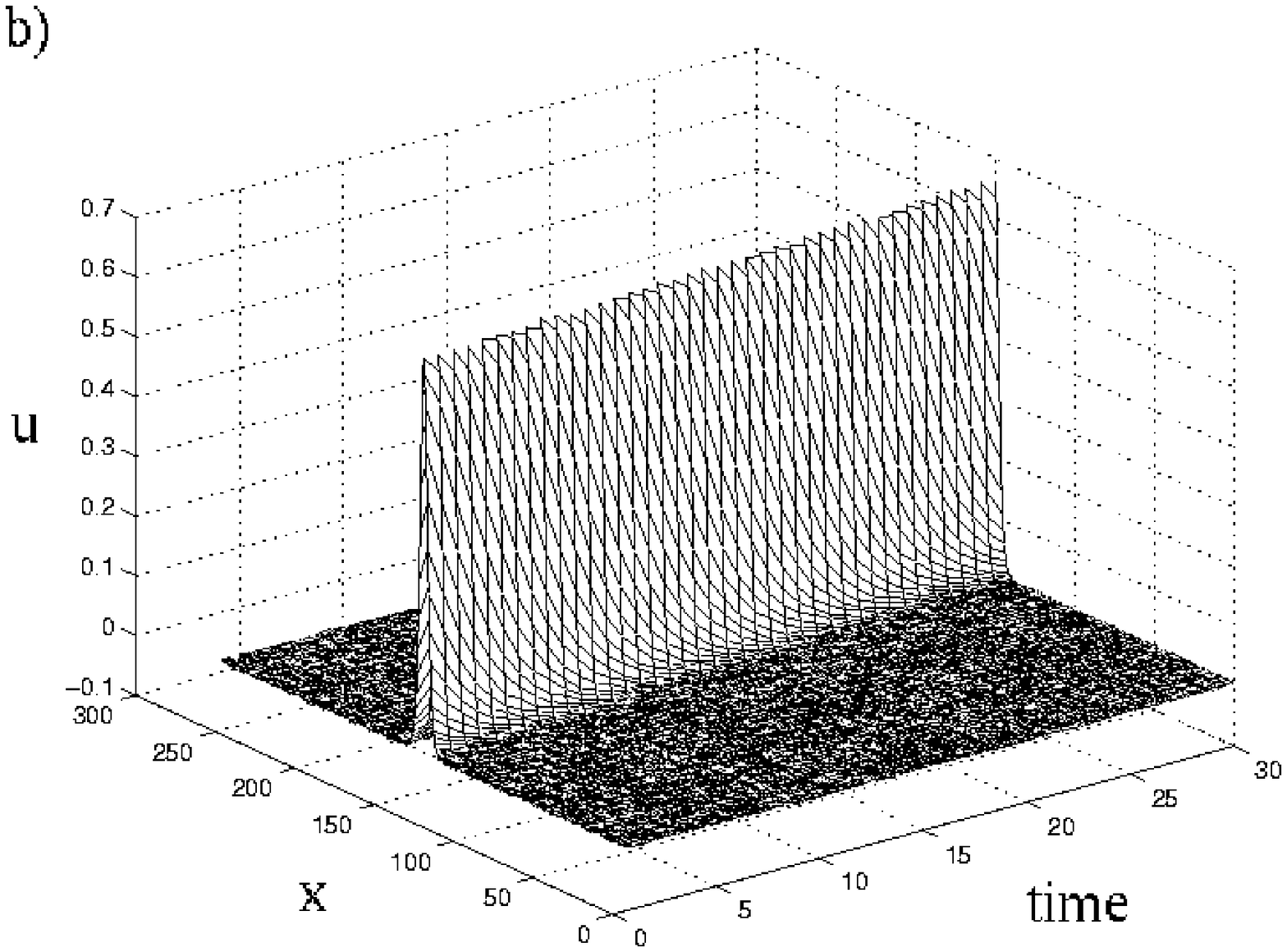}
\caption{a) Development of the longitudinal instability and collapse
  at time $t=12$ for $c=\frac{1}{4}$. b) Propagation of a stable
  solitary wave for $c=\frac{3}{4}$.}
\end{figure}
\begin{figure}
\epsfig{width=5cm,file=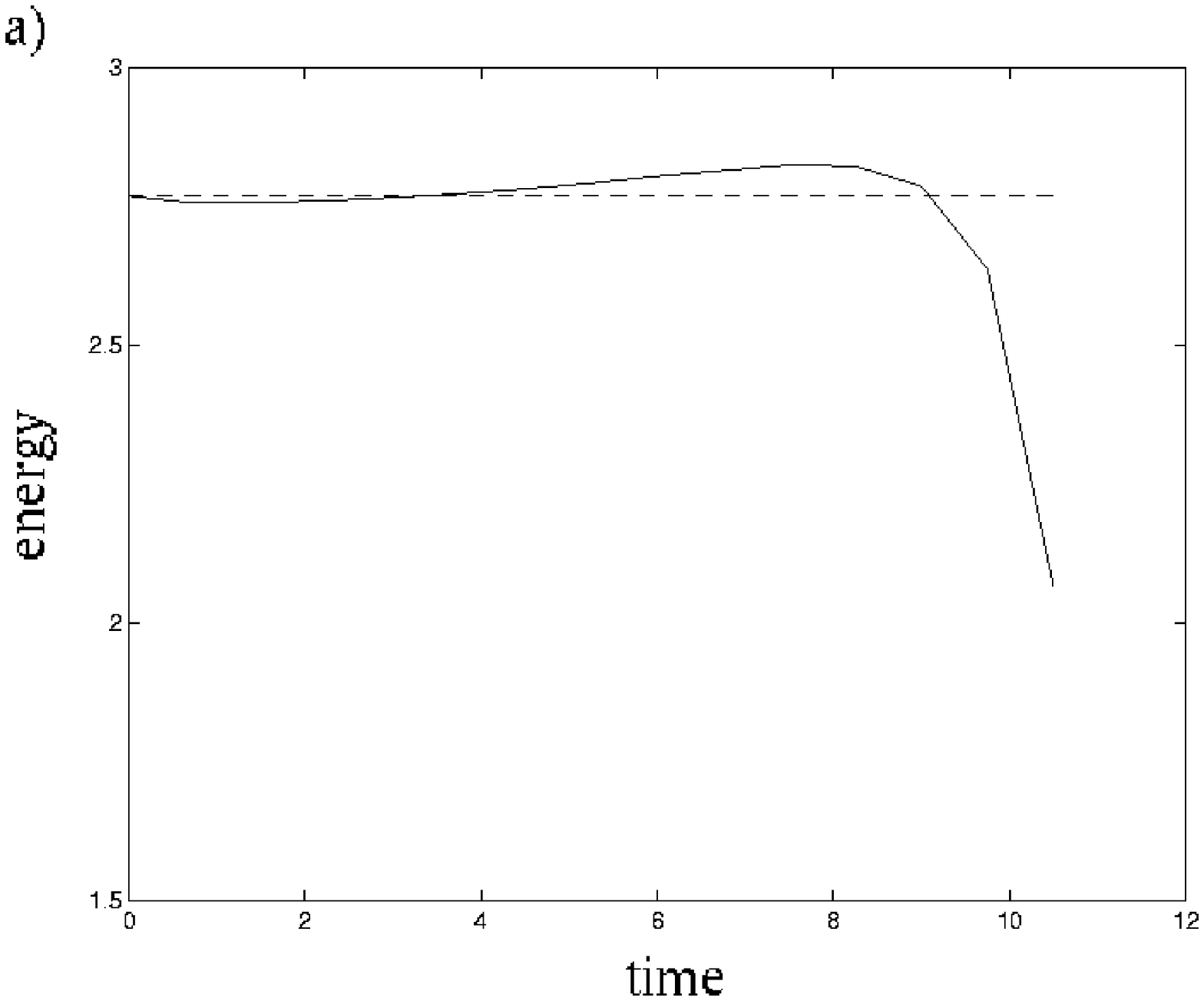}
\hspace{1cm}
\epsfig{width=5cm,file=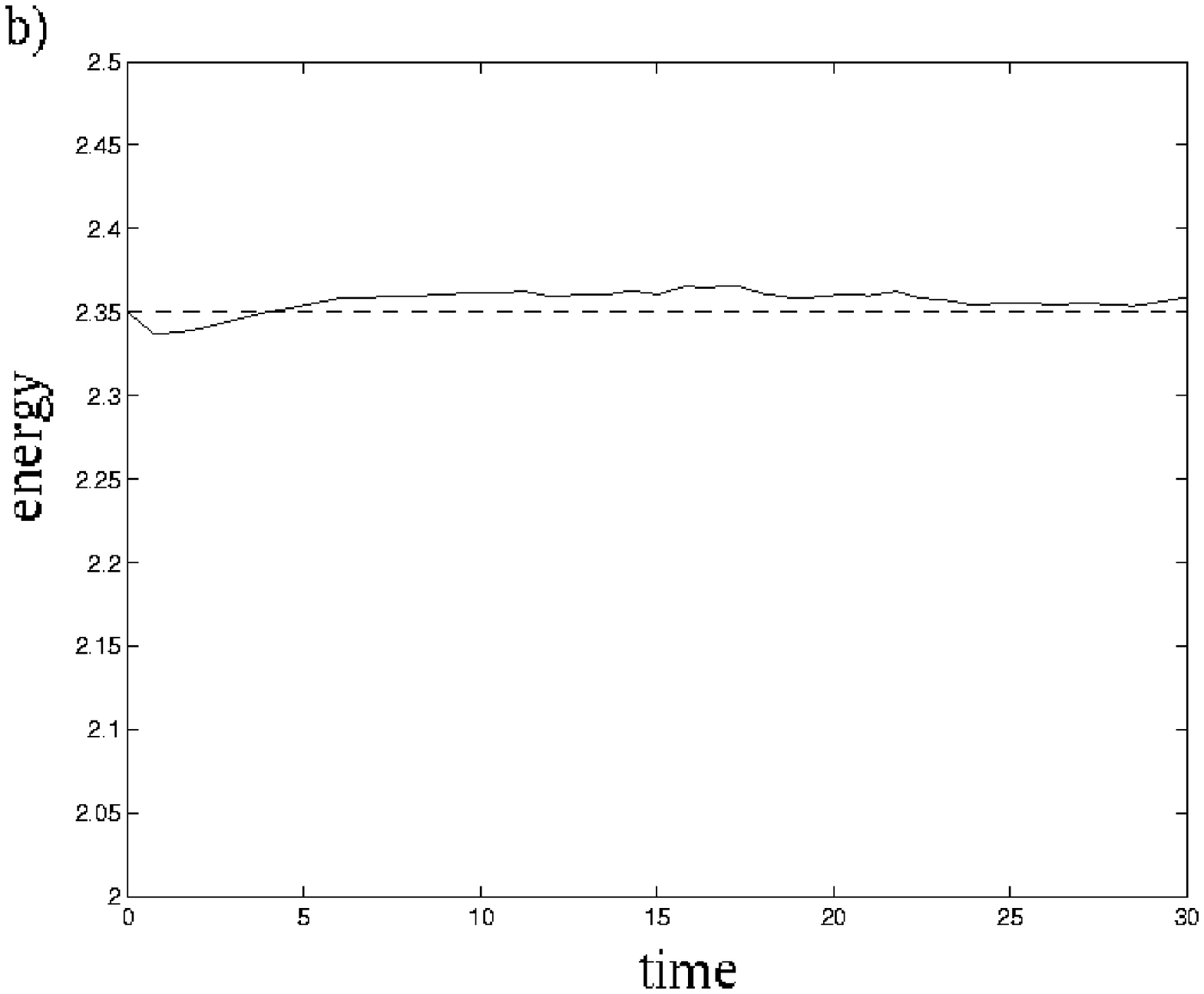}
\caption{Energy evolution. Dashed line represents the initial energy
  level, and the solid line shows the time evolution of
  energy. a) Unstable case $c=\frac{1}{4}$. b)
  Stable case $c=\frac{3}{4}$.}
\end{figure}
which follow from the periodicity of the discrete Fourier transform
\cite{F}, and $\bar{F}({\bf
  U})$ denotes the Fourier transform of the anti-derivative of the
function $f(u)$ in (\ref{1}). The same result would be obtained if one
applied the spectral discretization to the multi-symplectic
formulation (\ref{3}), as it was done for the Zakharov-Kuznetsov
equation in \cite{BR}.

For the second-order time derivative we used the central difference
approximation (time step was chosen to be $\Delta t=0.01$ in all the
simulations):
\begin{equation}
\partial_{t t}U_{k l}=\frac{U_{k l}^{n+1}-2U_{k l}^{n}+U_{k l}^{n-1}}{\Delta t^{2}}.
\end{equation}
One should note that the only valid test of this scheme can be
done for the ``good'' Boussinesq equation with $\sigma>0$. For
$\sigma<0$ in the case of the ``good'' Boussinesq equation, an initial
profile independent of $x$ would result in a solution which
could grow ``faster than exponential'' because for large transverse
wavenumbers, growth rate of the initial data has no upper bound
(ill-posedness).

To test the algorithm, we first used it to confirm the results for the
dynamics of the one-dimensional solitary waves. The initial profile
was taken to be of the form
\begin{equation}
u(x)=\frac{3}{2}\left(1-c^{2}\right)\sech^{2}\left[\frac{1}{2}\left(1-c^{2}\right)\left(x-\frac{L}{2}\right)\right]+\xi(x),
\end{equation}
where $\xi(x)$ is a small random perturbation. The results are presented
in Figures 3 and 4.  For $c=\frac{1}{4}$ the solitary wave solution is
linearly unstable as reported in \cite{AS,BD1}, and the development of
this linear instability is shown in Figure 3a). In the case
$c=\frac{3}{4}$ the numerical results confirm the stability of
the solitary wave (see Figure 3b) ). The
simulations were run on an interval of the length $L=256$ with
$2N=512$. As a numerical check, the total energy was monitored,
and it was found to be well behaved till near the collapse when the
significant errors occur, as illustrated in Figure 4.
\begin{figure}
\epsfig{width=5cm,file=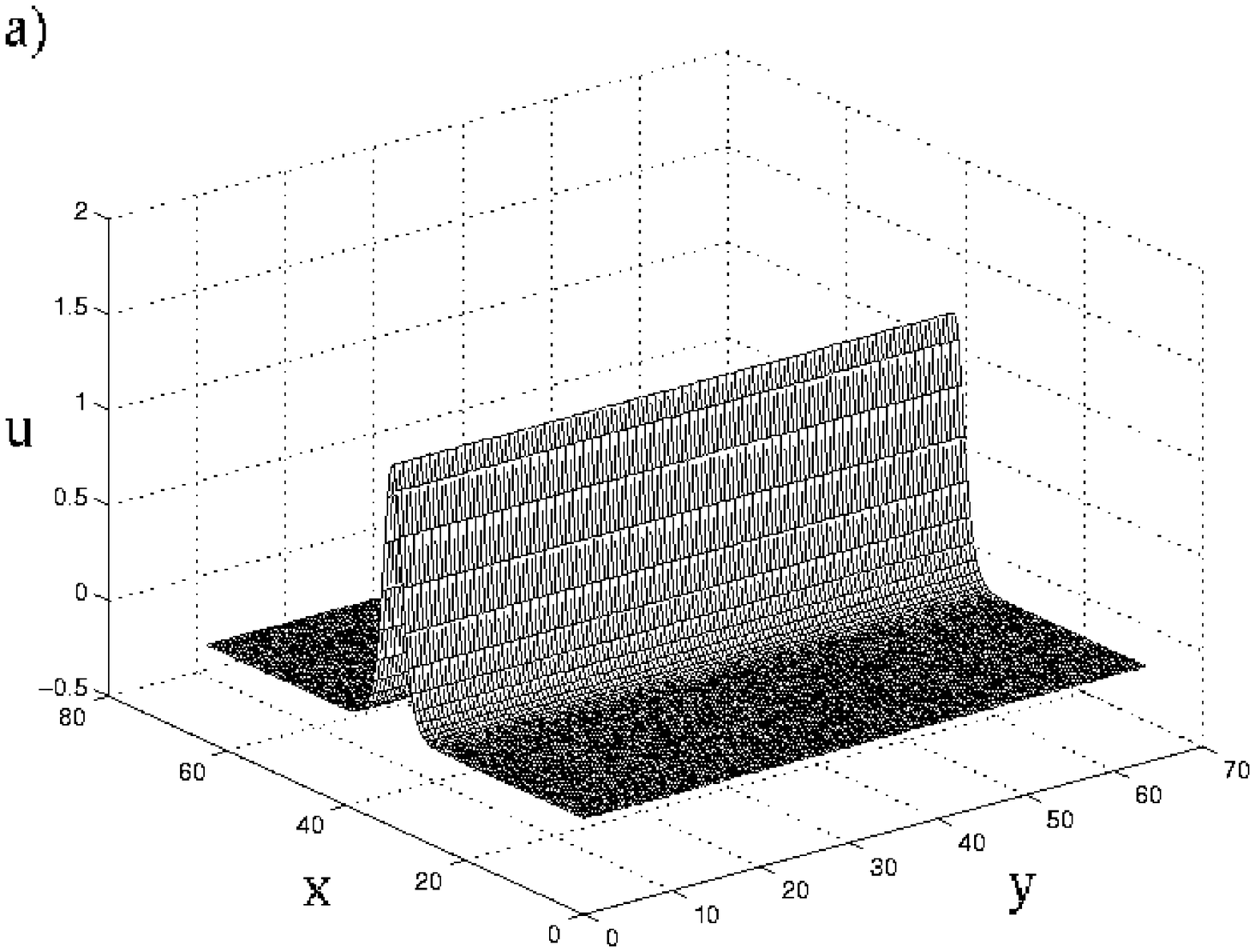}
\hspace{1cm}
\epsfig{width=5cm,file=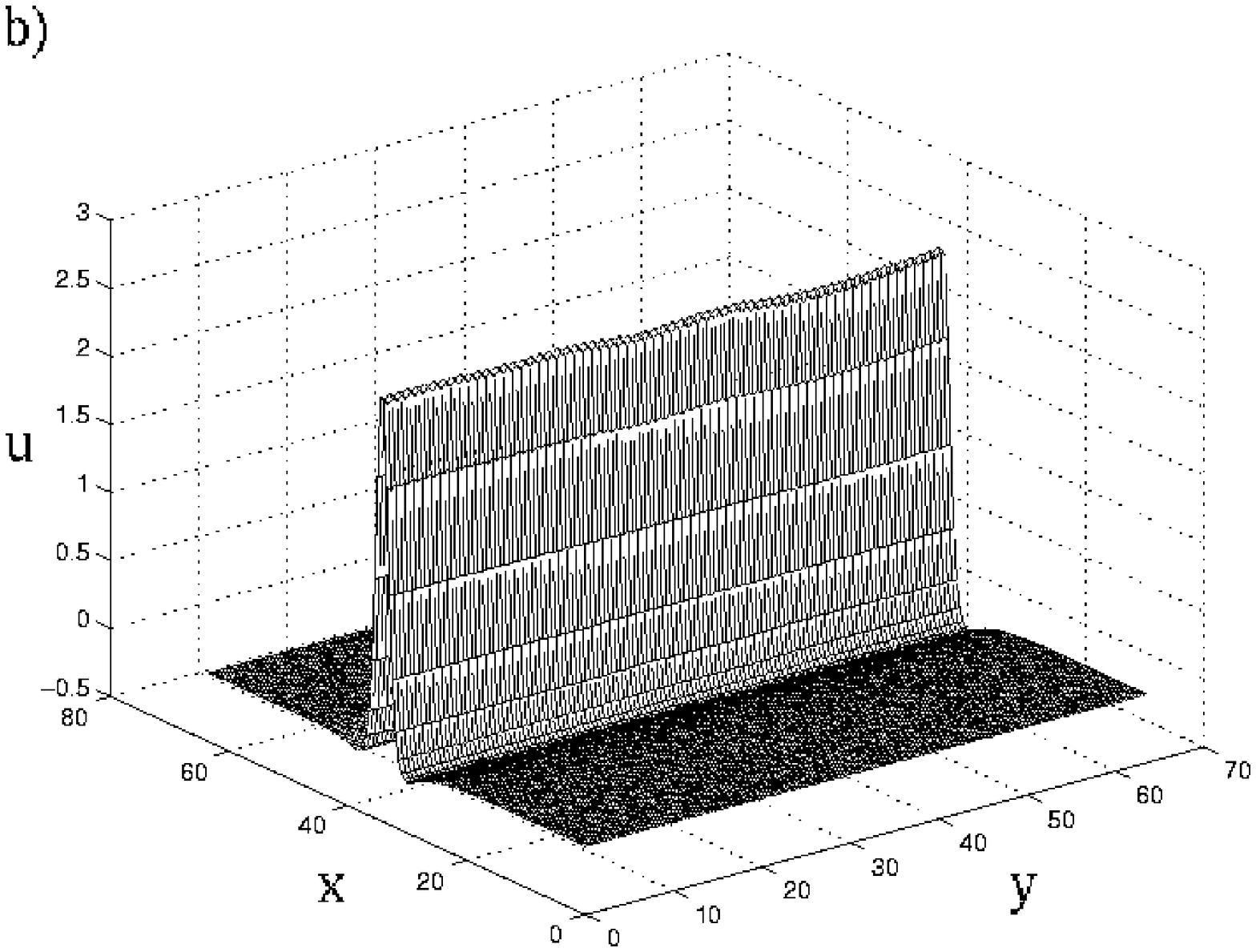}
\caption{ Solitary wave for $\sigma=1$ and $c=\frac{1}{4}$. a)
  Initial profile. b) Development of the transverse modulation (time
  $t=11.25$ ).}
\end{figure}
\begin{figure}
\epsfig{width=5cm,file=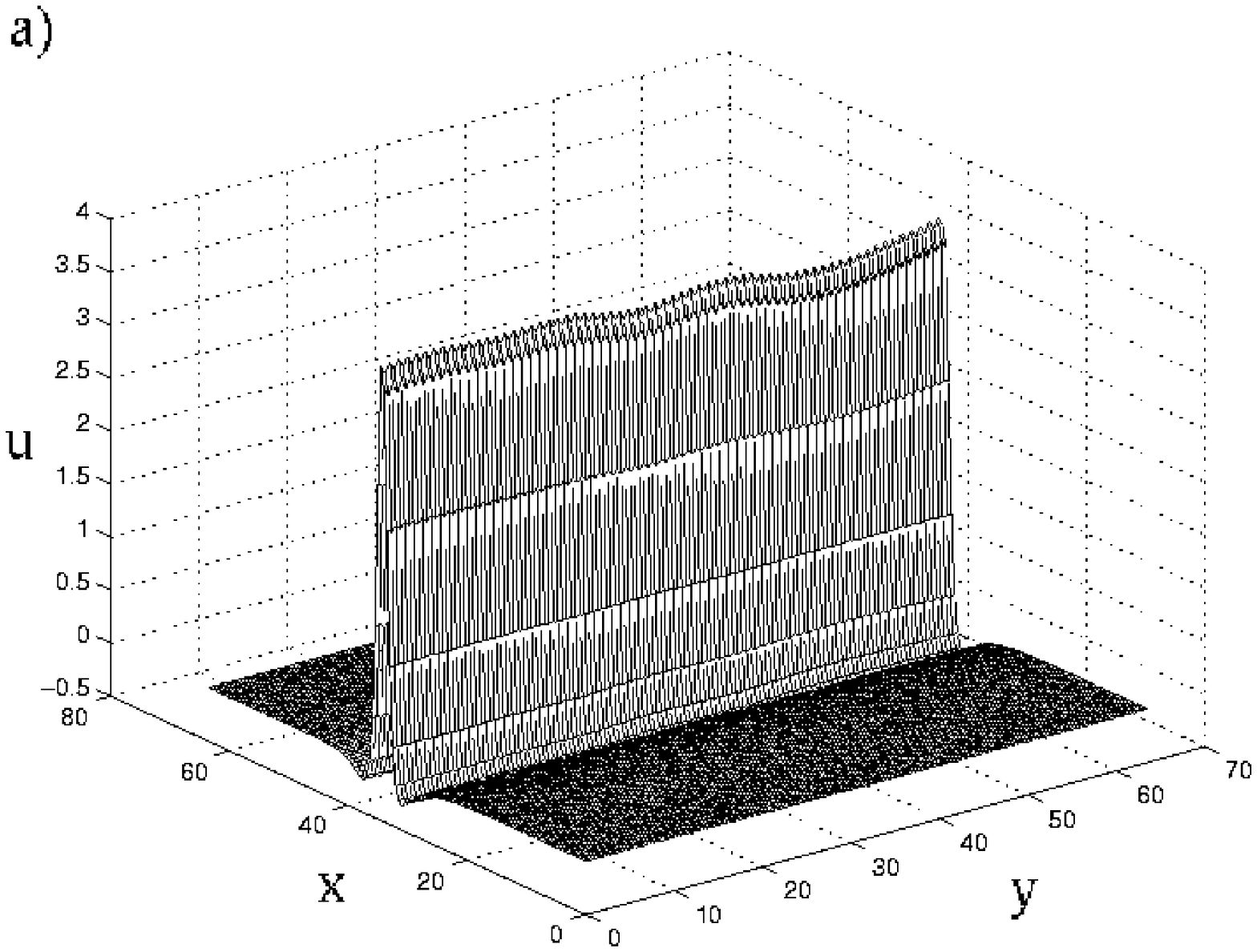}
\hspace{1cm}
\epsfig{width=4.7cm,file=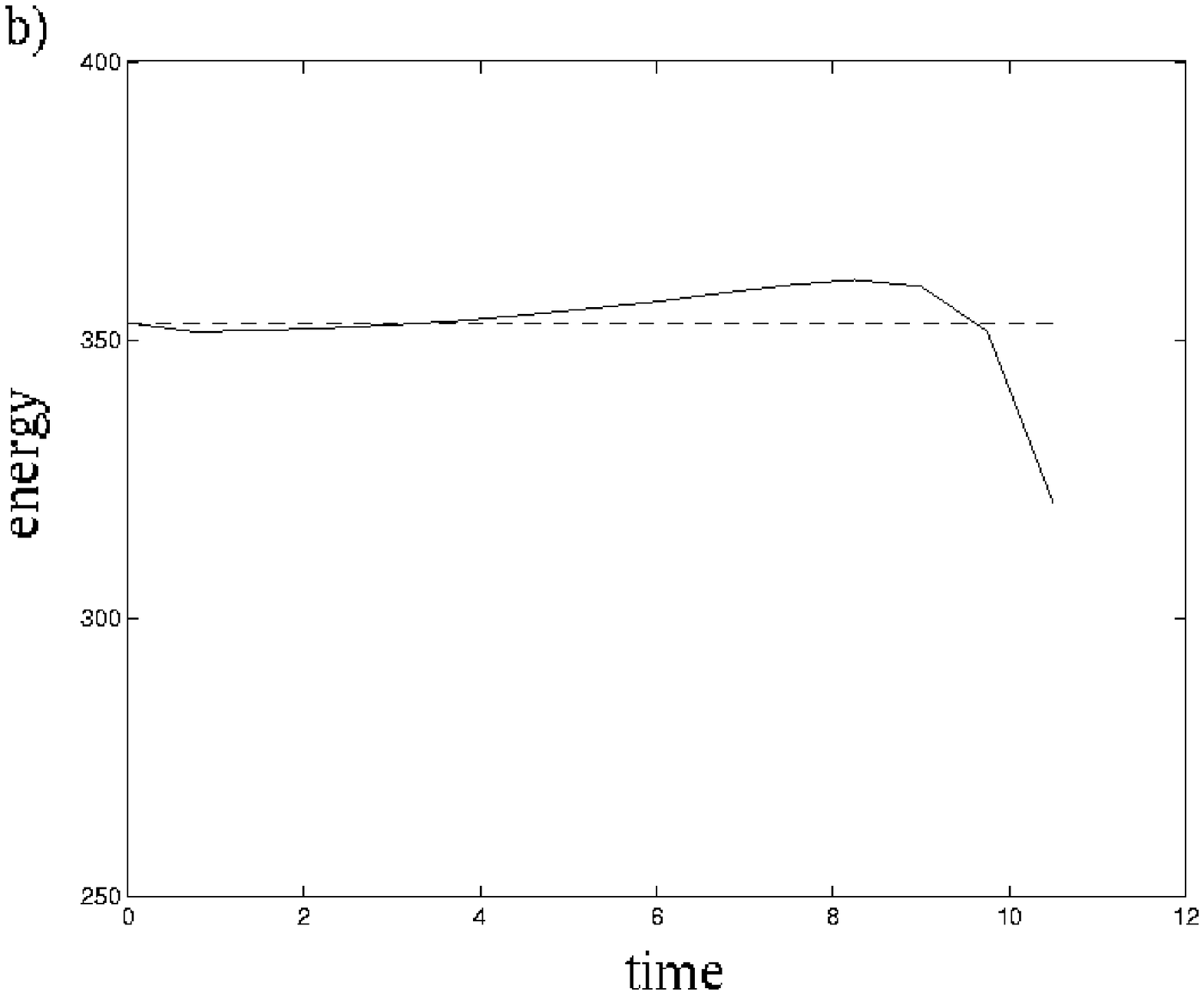}
\caption{The same as in Fig. 5. a) Wave collapse ( time $t=12$). b) Energy
  evolution. Dashed line represents the initial energy level, and the
  solid line shows the time evolution of  energy.}
\end{figure}

For the two-dimensional simulations we took an initial profile in the form
of the line solitary wave uniform in $y$
\begin{equation}\label{45}
u(x,y,0)=\frac{3}{2}\left(1-c^{2}\right)\sech^{2}\left[\frac{1}{2}\left(1-c^{2}\right)\left(x-\frac{L}{2}\right)\right]+\xi(x,y),
\end{equation}
\begin{figure}
\epsfig{width=5cm,file=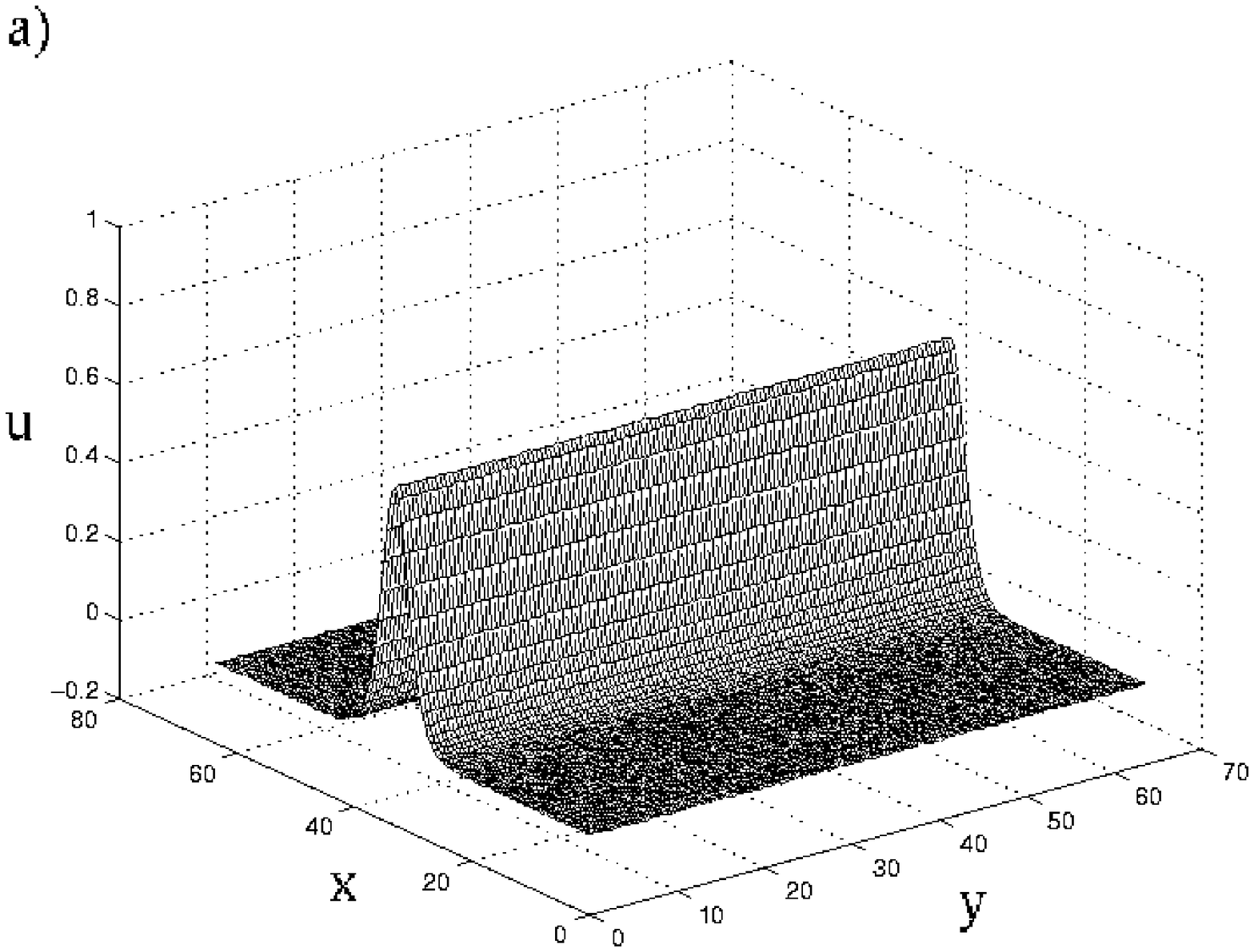}
\hspace{1cm}
\epsfig{width=5cm,file=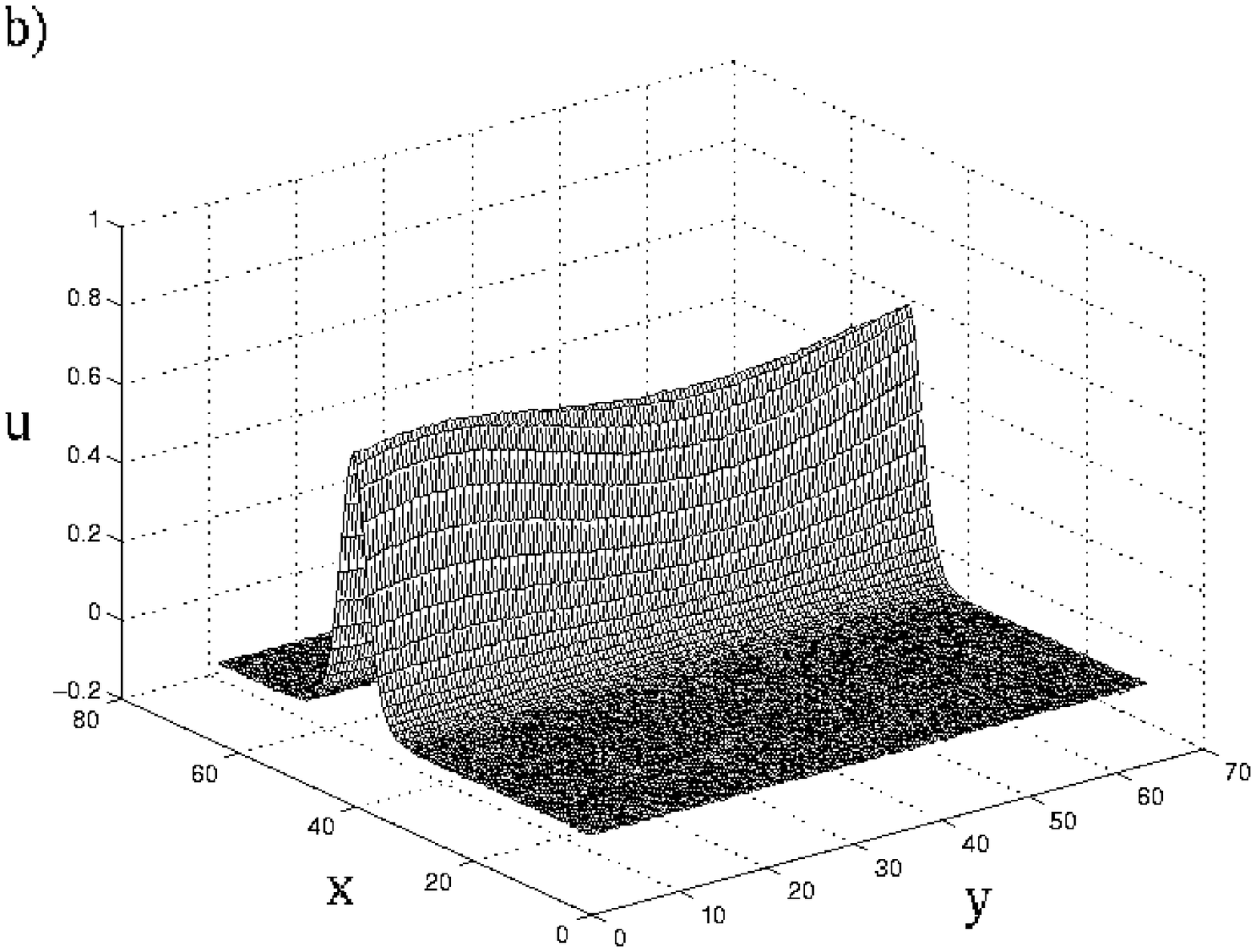}
\caption{ Solitary wave for $\sigma=1$ and $c=\frac{3}{4}$. a)
  Initial profile. b) Development of the transverse modulation (time
  $t=184.5$).}
\end{figure}
where $\xi(x,y)$ is a small random perturbation (in this case
$l=0$). The length of the square box was chosen to be $L=64$ with the
number of Fourier modes $2N=128$. In the case $c=\frac{1}{4}$ the
solitary wave (\ref{45}) is linearly unstable in longitudinal
direction as is known from the stability analysis of the 1D
equation. In Figure 5b) we can see this instability developing in a
similar way as in the 1D case. One can also note in this Figure the
development of the stable transverse modulation. Wave collapse in this
case is shown in Figure 6a), with the plot of energy as a function of
time in Figure 6b). When $c=\frac{3}{4}$, the solitary wave is
longitudinally stable but transversely unstable, and the development
of this instability is presented in Figures 7 and 8. We note that at the
initial stage of the evolution there is a transverse modulation
developing while the amplitude of the wave is gradually growing Figure
7b), then the instability prevails leading finally to the collapse of
the wave Figure 8a). Note that this collapse is clearly a
two-dimensional effect, since it does not happen uniformly in the
$y$-direction. The energy proves to be conserved rather well
during the simulations (see Figure 8b) ), although the energy deviates
substantially as the wave approaches the stage of collapse.
\begin{figure}
\epsfig{width=5cm,file=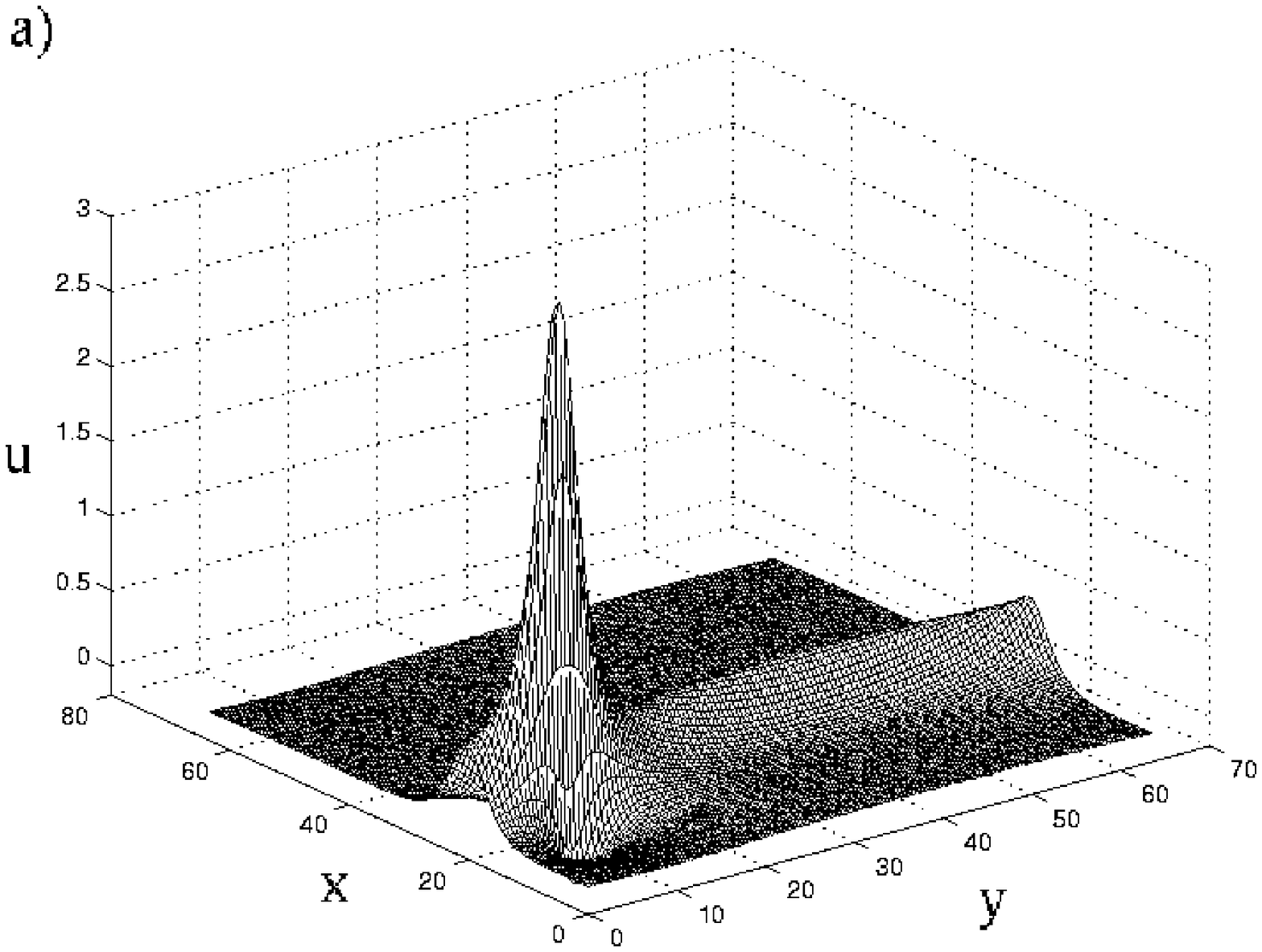}
\hspace{1cm}
\epsfig{width=4.7cm,file=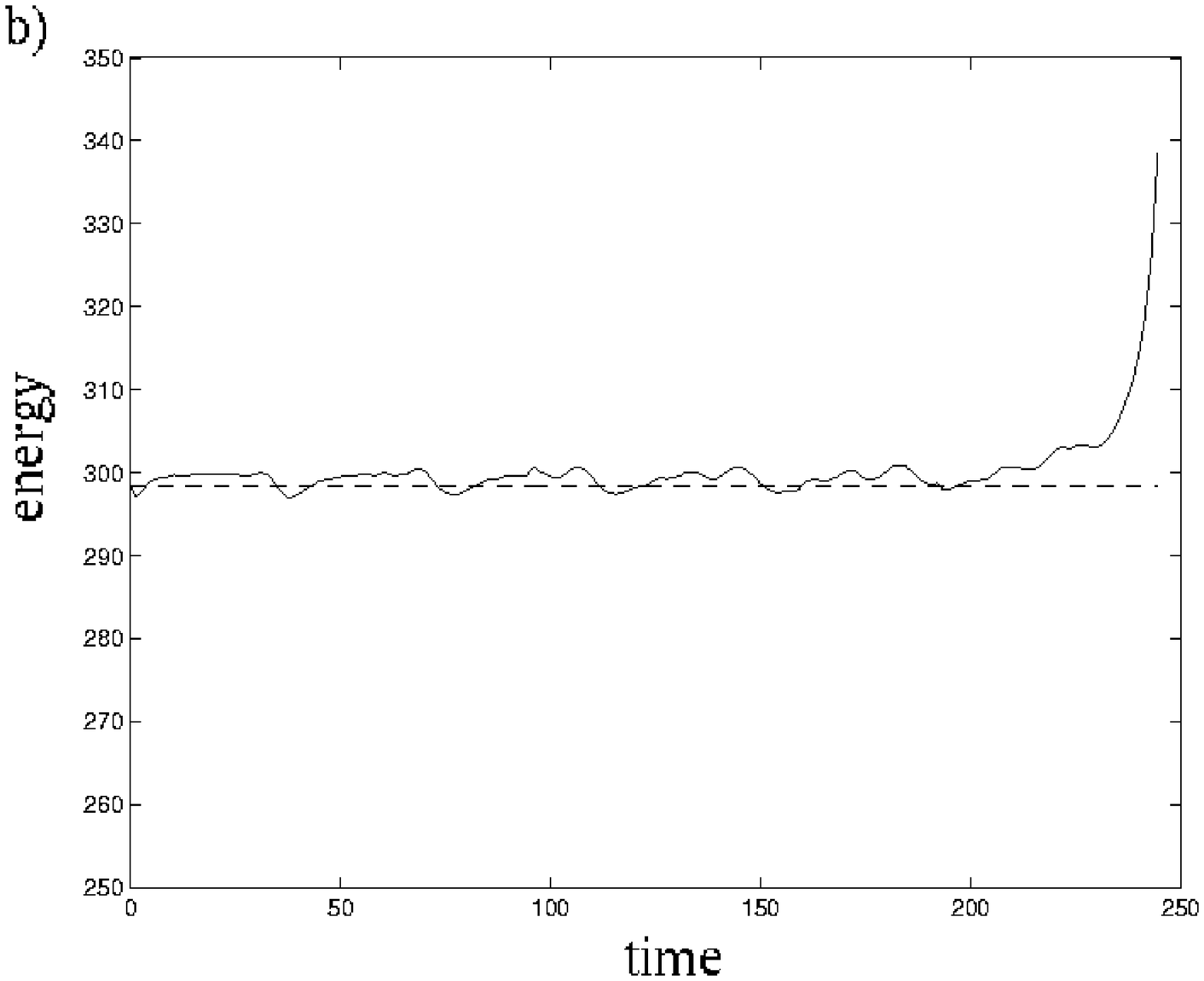}
\caption{The same as in Fig. 7. a) Wave collapse (time $t=246$). b) Energy
  evolution. Dashed line represents the initial energy level, and the
  solid line shows the time evolution of energy.}
\end{figure}

\section{Concluding remarks}

We have considered the transverse instability of line solitary wave
solutions of the (2+1)-dimensional Boussinesq equation. Using the
multi-symplectic formulation of the system, we derived a
geometric condition for this instability for small transverse
wavenumbers. With an Evans function approach, the linearised stability
equation was analyzed, and this allowed to obtain the dependence of
the instability growth rate for all transverse wavenumbers. Numerical
simulations support the analytical results about transverse and
longitudinal instabilities and demonstrate the development of those
instabilities and subsequent wave collapse.  

We conclude with an open problem.
While analytic theories for collapse of solitary waves for the
Boussinesq equation in one space dimension exist \cite{T}, it is an
interesting open problem to develop an analytical technique for
predicting collapse for the case of two space dimensions, e.g. a
generalization of the virial theorem or the result of \cite{T} for
example, and moreover, to determine if transverse instability for
(\ref{1}) leads to collapse for {\it all\/} parameter values.

\section{Acknowledgements}

The authors would like to thank Sebastian Reich for advice on the
numerics and Georg Gottwald for helpful discussions.


\begin{thebibliography}{50}

\bibitem{AGJ} J.C. Alexander, R. Gardner, C.K.R.T. Jones, J. Reine
  Angew. Math. {\bf 410}, 167 (1990).

\bibitem{APS} J.C. Alexander, R.L. Pego, R.L. Sachs, Phys. Lett. A
  {\bf 226}, 187 (1997).

\bibitem{AS} J.C. Alexander, R.L. Sachs, Nonlinear World {\bf 2}, 471 (1995).

\bibitem{AR} M.A. Allen, G. Rowlands, J. Plasma Phys. {\bf 50}, 413 (1993).

\bibitem{ARa} M.A. Allen, G. Rowlands, Phys. Lett. A {\bf 235}, 145 (1997).

\bibitem{B} J.G. Berryman, Phys. Fluids {\bf 19}, 771 (1976).

\bibitem{Web} K.B. Blyuss,
{\tt http://www.maths.surrey.ac.uk/personal/pg/K.Blyuss/}

\bibitem{BM} B.N. Breizman, V.M. Malkin, Sov. Phys. JETP {\bf 52}, 435 (1980).

\bibitem{BS} J.L. Bona, R.L. Sachs, Comm. Math. Phys. {\bf 118}, 15 (1988).

\bibitem{B1} T.J. Bridges, Phys. Rev. Lett. {\bf 84}, 2614 (2000).

\bibitem{B3} T.J. Bridges, J. Fluid Mech. {\bf 439}, 255 (2001).

\bibitem{BD2} T.J. Bridges, G. Derks, Phys. Lett. A {\bf 251}, 363 (1999).

\bibitem{BD1} T.J. Bridges, G. Derks, Arch. Rat. Mech. Anal. {\bf
    156}, 1 (2001).

\bibitem{BD3} T.J. Bridges, G. Derks, University of Surrey Preprint,
  2001 (to be published).

\bibitem{BR} T.J. Bridges, S. Reich, Physica D {\bf 152-153}, 491 (2001).

\bibitem{F} B. Fornberg, {\it A practical guide to pseudospectral methods}
  (Cambridge University Press, Cambridge, 1996).

\bibitem{IRS} E. Infeld, G. Rowlands, A. Senatorski,
  Proc. Roy. Soc. Lond. A {\bf 455}, 4363 (1999).

\bibitem{J} R.S. Johnson, J. Fluid Mech. {\bf 323}, 65 (1996).

\bibitem{KP} B.B. Kadomtsev, V.I. Petviashvili, Sov. Phys. Dokl. {\bf
    15}, 539 (1970).

\bibitem{KPe} Y.S. Kivshar, D.E. Pelinovsky, Phys. Rep. {\bf
  331}, 117 (2000).

\bibitem{KRZ} E.A. Kuznetsov, A.M. Rubenchik, V.E. Zakharov,
  Phys. Rep. {\bf 142}, 104 (1986).

\bibitem{LS} E.W. Laedke, K.H. Spatschek, Phys. Rev. Lett. {\bf 41},
  1798 (1978).

\bibitem{T} S.K. Turitsyn, Phys. Rev. E {\bf 47}, R796 (1993).

\bibitem{WAS} X.P. Wang, M.J. Ablowitz, H. Segur, Physica D {\bf 78},
  241 (1994).

\bibitem{VEZ} V.E. Zakharov, Sov. Phys. JETP {\bf 26}, 994 (1968) 994.

\end{thebibliography}
\end{document}